\documentclass[twocolumn,showpacs,preprintnumbers,amsmath,amssymb,floatfix]{revtex4}

\usepackage{graphicx}
\begin{document}

\title{Coherence scale of the two-dimensional  Kondo Lattice model.}
\author{F.F. Assaad}
\affiliation{ Institut f\"ur Theoretische Physik und Astrophysik
Universit\"at W\"urzburg, Am Hubland D-97074 W\"urzburg }

\begin{abstract}
A doped hole in the two-dimensional half-filled Kondo lattice model with exchange $J$ and 
hopping $t$ has momentum $(\pi,\pi)$ irrespective  of the coupling $J/t$.  The quasiparticle 
residue of the doped hole,  $Z_{ \left( \pi, \pi \right) } $,  tracks the Kondo scale, $T_K$, 
of the corresponding single impurity model. 
Those results stem from high precision  quantum Monte Carlo simulations on lattices up to $12 \times 12$.
Accounting for small  dopings away from half-filling within a rigid band 
approximation,  this  result implies that  the effective mass of the  charge carriers at the Fermi 
level tracks $1/T_K$ or equivalently that the coherence  temperature $T_{\rm coh} \propto T_K$. 
This results is consistent with the large-N saddle point of the  $SU(N)$ symmetric Kondo 
lattice model.   
\end{abstract}

\maketitle

In a Fermi liquid, the coherence scale corresponds to the energy scale below which thermodynamic
properties are determined by the Fermi surface:  the specific heat is linear in temperature and
the spin charge uniform susceptibilities temperature independent. This  scale is 
inversely proportional to the  effective mass of the charge carriers. In this article, our aim 
is to compute the coherence scale in models of heavy fermion materials \cite{Lee86}. 
Our starting point is the  Kondo Lattice model (KLM),
\begin{equation}
\label{KLM}
	H = -t\sum_{\langle { \vec{i}, \vec{j} } \rangle, \sigma } [ c^{\dagger}_{\vec{i},\sigma } 
c_{\vec{j},\sigma }  + {\rm h.c.} ]  + J \sum_{\vec{i}} \vec{S}^{c}_{\vec{i}} \vec{S}^{f}_{\vec{i}},
\end{equation}
on a square lattice. 
The first term  describes a  band of non-interacting electrons in a tight binding approximation 
($c^{\dagger}_{\vec{i},\sigma }$ creates an electron in a Wannier state centered around lattice site 
$\vec{i}$  with z-component of spin $\sigma$) and the sum runs over nearest neighbors. 
The spin degrees of freedom of the conduction electrons, 
$ \vec{S}^{c}_{\vec{i}} = \frac{1}{2} \sum_{s,s'} c^{\dagger}_{\vec{i},s} \vec{\sigma}_{s,s'} 
 c^{\dagger}_{\vec{i},s'}$ with $\vec{\sigma}$ the Pauli spin matrices, interact antiferromagnetically,
$J>0$,  with a lattice of spin $1/2$ magnetic impurities, 
$\vec{S}^{f}_{\vec{i}}$.  The KLM at $J/t << 1$ stems from the 
periodic Anderson model (PAM) in the limit where charge fluctuations are negligible. 

In  the limit of a single impurity, the model is well understood: 
at {\it high} temperatures the impurity spin is   essentially free  and at low temperatures it 
is screened via the formation of a many body singlet state of the conduction electrons and 
impurity spin. The 
characteristic, universal,  energy scale  describing this crossover from the  free to the screened 
magnetic impurity is the Kondo temperature, $T_K$ \cite{Hewson}.  
In the lattice case, it is appealing to view the heavy fermion metallic state
as a consequence of a coherent, Bloch-like,  superposition of individual Kondo screening clouds.
The coherence  scale of the metallic state has been investigated in details within 
the large-N approximation for the KLM \cite{Georges00} and the dynamical mean field approximation for 
the  related  PAM  \cite{Pruschke00}.  Both  approaches yield  $T_{\rm coh} \propto T_K$ as 
a function of $J/t$  with a proportionality factor  depending strongly on the band-filling.   
Since both approximations neglect spatial  fluctuations, they do not capture the 
Ruderman-Kittel-Kasuya-Yosida (RKKY)  interaction
which introduces a new energy scale into the problem and drives the ground state through a 
magnetic order-disorder quantum phase transition.  The new result in this Letter 
is that the relation $T_{\rm coh} \propto T_K$  still holds 
when the  two-dimensional  KLM is solved numerically  exactly 
by means of quantum Monte Carlo (QMC)  simulations. 

To tackle this problem, we have to adopt 
an indirect route  since the sign problem inhibits  simulations away from half-filling. 
We hence investigate the problem of a  single-hole doped  into the Kondo insulating state at 
half-filling and then assume a rigid band to deduce the properties of the metallic state   at weak 
dopings. 
In the following, we describe our approach  for the $SU(N)$ symmetric model where in the limit 
$N \rightarrow \infty $  a paramagnetic saddle point approximation becomes exact. We then compare 
the obtained results with QMC simulations for the $SU(2)$  model of Eq. \ref{KLM}.

{\it Mean Field.}   The mean field we consider neglects magnetic fluctuations triggered 
by the RKKY interaction  but  describes 
the Kondo effect as well as the formation of the heavy electron state. The decoupling 
is based on the equation: $ \vec{S}^{c}_{\vec{i}} \vec{S}^{f}_{\vec{i}} = \frac{1}{4} 
\left( D^{\dagger}_{\vec{i}} D_{\vec{i}} + D_{\vec{i}} D^{\dagger}_{\vec{i}} \right)$ with 
$D_{\vec{i}} = \sum_{\sigma} c^{\dagger}_{\vec{i},\sigma } f_{\vec{i},\sigma }$. Here,
$ \vec{S}^{f}_{\vec{i}} = \frac{1}{2} \sum_{s,s'} f^{\dagger}_{\vec{i},s} \vec{\sigma}_{s,s'} 
 f^{\dagger}_{\vec{i},s'}$, and  $f^{\dagger}_{\vec{i},\sigma} $ are fermionic operators.
 At the  mean-field
level one  adopts a real  order parameter,  
$ r= \langle D_{\vec{i}} \rangle$, and the  constraint of a single charge per f-site  is imposed on average 
via a Lagrange multiplier $\lambda$. This mean field theory has been discussed extensively 
in \cite{Georges00}.  Here, we will concentrate on the $J$ dependence of scales 
for a half-filled particle-hole symmetric conduction band. This symmetry pins
 the chemical potential to $\mu = 0$ and $\lambda=0$. 
Single particle properties are derived from  the Green function 
$G(\vec{k},i\omega_m) = 1/ \left[ i \omega_m - \epsilon(\vec{k}) - \Sigma( i \omega_m) \right] $ 
with self-energy
\begin{equation}
\label{Self}
         \Sigma(i\omega_m) = \frac{\left(Jr/2\right)^2}{ i \omega_m}
\end{equation}
and $\epsilon(\vec{k}) = -2t \left( \cos(k_x) +  \cos(k_y) \right)$.
Assuming a  flat density of states, 
$D(\epsilon)$, of width $W$  satisfying $\int {\rm d } \epsilon D(\epsilon) = 1$ and
$D(\epsilon) = D(-\epsilon)$  we can solve the saddle point  equations to obtain: 
\begin{equation}
	r(T=0) = \frac{W}{2J \sinh(W/2J)}. 
\end{equation}
The order parameter $r$ vanishes at the Kondo temperature $T_K$. At the mean-field level, this 
energy scale is independent on the impurity concentration and  hence  matches the 
Kondo temperature of the corresponding single impurity problem. 
With the  above flat density of states we obtain: 
\begin{equation}
	T_K \propto W e^{-W/J} \propto \frac{1}{W}\left[ J r(T=0) \right]^2  \; {\rm for} \;  J/t << 1. 
\end{equation}

The functional form of the self-energy leads to the dispersion relation
$ E_{\vec{k}}  = \frac{1}{2} \left[\epsilon(\vec{k}) - \sqrt{ \epsilon^2(\vec{k})+ (Jr)^2} \right]  $ with 
residue $ Z_{\vec{k}}  =  \frac{1}{2} 
\left[ 1 - \epsilon(\vec{k}) / \sqrt{\epsilon^2(\vec{k})+ (Jr)^2 } \right] $ for a doped hole away 
from half-filling.  At $T=0$ and in the weak-coupling limit, both the quasiparticle gap, 
$\Delta_{qp} = \mu - E_{\pi,\pi}$,  and residue track the same scale: 
\begin{equation}
	\Delta_{qp} \propto Z_{(\pi,\pi)} \propto [Jr(T=0)]^2 \propto T_K \; {\rm for} J/t << 1.
\end{equation}

Under  a rigid band assumption (see Discussion section)  this results implies that the 
coherence temperature of the metallic state  at small hole dopings away from half-filling 
tracks the  the Kondo scale. At the mean-field level this statement may be checked explicitly by solving the 
mean-field equations in the metallic state \cite{Georges00}.

The mean-field theory has many caveats. The  order parameter $r$ does not 
vanish  because the constraint is taken into account on average.  Magnetism is 
which is energetically favorable in the small $J/t$ limit is not included. Hence it is a-priori 
not clear that the scales described by the mean-field approach  survive when  the model is solved exactly. 

\begin{figure}[h]
\begin{center}
\includegraphics[width=.45\textwidth]{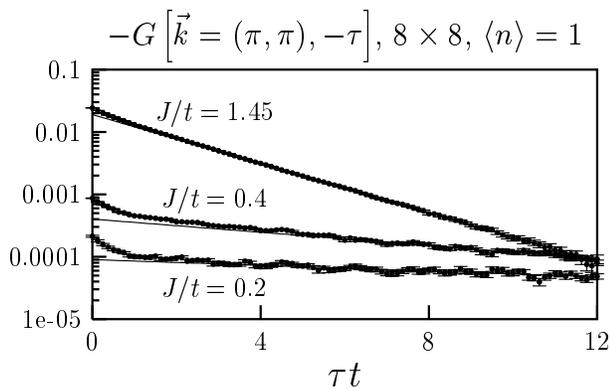} \\
\end{center}
\caption[]{ Zero temperature  Green function defined in Eq. \ref{Green.eq}. Fitting the tail of 
this quantity to  the form  $Z_{\vec{k}}e^{-\Delta_{qp}(\vec{k}) \tau}$ yields the quasiparticle residue 
and quasiparticle gap at wave vector $\vec{k}$ } 
\label{Green.fig}
\end{figure} 

{\it QMC.} To obtain zero temperature properties, we solve the half-filled KLM with 
the projector  auxiliary field QMC (PQMC) method. The application of the method to the KLM has  
been describe extensively in  Ref. \cite{Assaad99a,Capponi00}.  The algorithm scales as 
$\beta N^3$ where, in the framework of the PQMC, $\beta$ is an projection parameter in imaginary 
time and $N$ is the number of  orbitals.  Hence, large values of $\beta $ may be used to guarantee 
convergence to the ground state, but on the other hand the $N^3$ scaling renders large lattice sizes 
difficult to access.
From the imaginary time  displaced Green 
function, we  compute both the quasiparticle gap as well as quasiparticle residue:
\begin{equation} 
\label{Green.eq}
	-G(\vec{k},-\tau) = 
\langle \Psi_0^N | c^{\dagger}_{\vec{k},\sigma}(\tau) c_{\vec{k}, \sigma}(0) | \Psi_0^N \rangle  
\stackrel{\tau \rightarrow \infty}{\rightarrow}  Z_{\vec{k}}e^{-\Delta_{qp}(\vec{k}) \tau }  
\end{equation}
since  $Z_{\vec{k}} = \left|\langle \Psi_0^N | c^{\dagger}_{\vec{k},\sigma} | \Psi_0^{N-1} \rangle  \right|^{2}$.  Clearly high precision results are required to extract the residue. Fig. \ref{Green.fig} plots typical 
raw data. To reach such precision   we use a newly developed  method for the 
calculation of $G(\vec{k},\tau)$ within the PQMC \cite{Feldbach00}.

\begin{figure}[]
\begin{center}
\includegraphics[width=.45\textwidth]{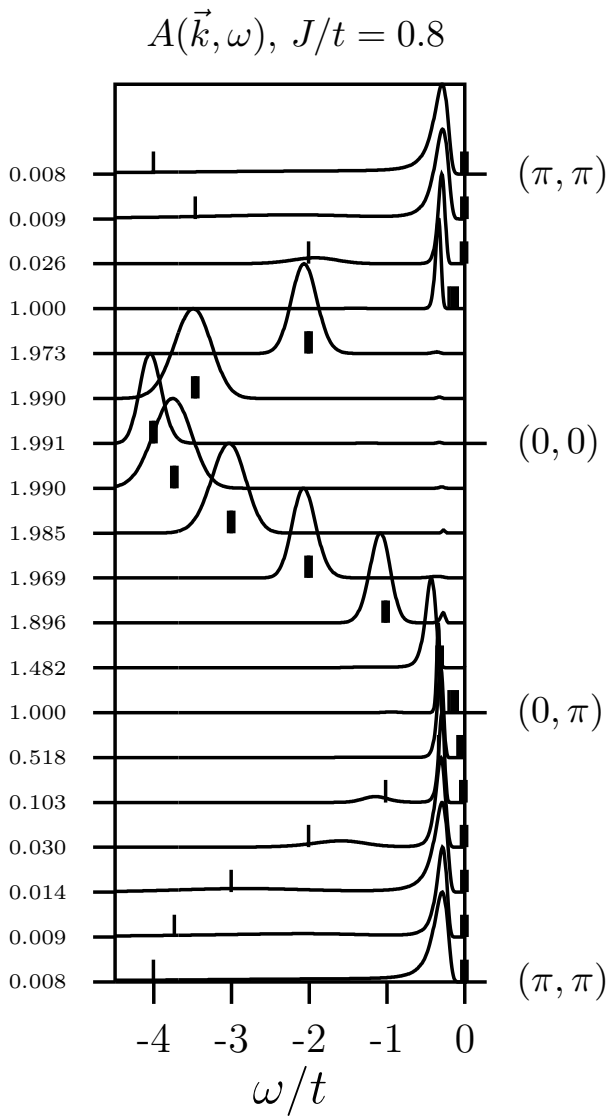} 
\end{center}
\caption[]{Single particle spectral function as obtained from analytically continuing the 
imaginary time QMC data with the Maximum Entropy method. The single  particle occupation 
numbers,  $n_{\vec{k} } =  \sum_{\sigma}
 \langle c^{\dagger}_{\vec{k},\sigma} c^{}_{\vec{k},\sigma} \rangle $ are listed 
on the left hand side of the plot, and correspond to the weight under each spectrum: 
$\pi n_{\vec{k}} = \int_{-\infty}^{0} {\rm d} \omega A(\vec{k},\omega)  $. The thin vertical lines 
track the dispersion relation of hole in a static staggered magnetic field  (see text): 
$   - \sqrt{ \epsilon^2(\vec{k}) + (J /4)^2 }  $. The bold vertical lines 
 correspond to the dispersion relation
stemming from the self-energy of Eq. \ref{Self}, with the order parameter $r$ determined self-consistently.}
\label{Akom.fig}
\end{figure} 

\begin{figure}[]
\begin{center}
\includegraphics[width=.45\textwidth]{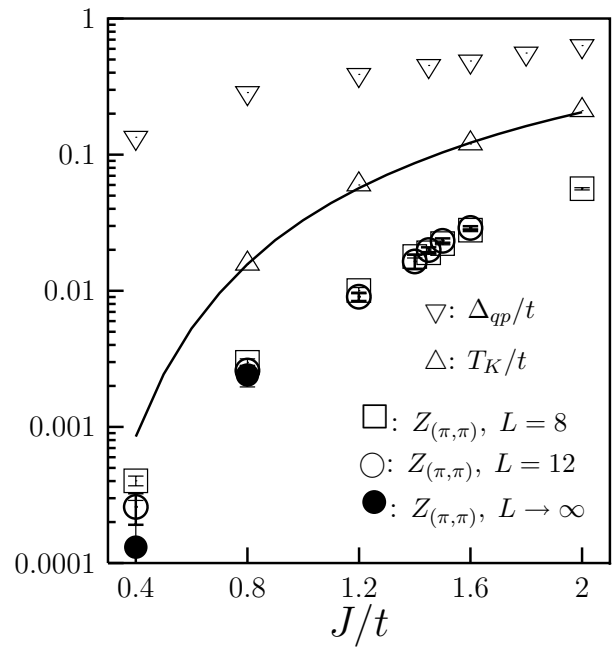} \\
\end{center}
\caption[]{ Scales as a function of $J/t$ as determined from QMC simulations (see text).     
The data for the quasiparticle gap are a result of extrapolation to the thermodynamic limit. The solid 
line corresponds to the mean-field (see section Mean-field) value of the Kondo temperature. Within our
resolution we cannot distinguish it from the QMC value. } 
\label{Scales.fig}
\end{figure}

{\it Numerical Results.}  In numerical simulations of the two-dimensional KLM,
the RKKY interaction   triggers a magnetic 
order-disorder quantum phase transition at  $J_c/t = 1.45 \pm 0.05 $ \cite{Capponi00}.  
The question then arises 
if the antiferromagnetic ordering destroys  Kondo screening and hence the possible 
appearance of a Kondo scale in the numerical data.
We  answer this question by analyzing the single particle spectral 
function.  Assuming that well under $J_c/t$  the impurity spins are frozen into their 
antiferromagnetic ordering, we can use the mean-field Ansatz, 
$ \langle S^{f,z}_{\vec{i}} \rangle = \frac{1}{2} m^f_z e^{i \vec{Q} \cdot \vec{i}} $, which 
leads to a single hole dispersion relation: 
$E(\vec{k}) = - \sqrt{ \epsilon^2(\vec{k}) + (J /4)^2 } $ \cite{Tsunetsugu97_rev}.  
 This approximation correctly reproduces 
the quasiparticle  gap $\Delta_{qp}  \propto J/4 $  in the small $J$ limit (see Fig. \ref{Scales.fig}) 
but fails to account for the observed dispersion relation in the spectral function.  
This statement is
based on  comparison with QMC data: the thin vertical lines in Fig. \ref{Akom.fig}, 
tracking the  dispersion  relation $   - \sqrt{ \epsilon^2(\vec{k}) + (J /4)^2 }  $, do  not account for 
the low energy features  around the $ (\pi,\pi)$ point. 
On the other hand, comparison with the mean-field approximation of the previous section (bold vertical 
lines in Fig. \ref{Akom.fig}), allows the interpretation that the low-lying features 
in the vicinity of  $\vec{k} = (\pi,\pi)$ stem from Kondo screening  and  are shifted
to  energy scales of order $J/t$ due to the  onset of magnetic ordering.  
We have checked that this functional 
form of the dispersion relation survives  down to $J/t = 0.2$ on $8 \times 8$ lattices.  
In the magnetically ordered phase, a doped 
hole at  $\vec{k} = (\pi,\pi)$ can scatter off a gapless magnon with momentum $\vec{Q} = (\pi,\pi)$ thus 
generating a shadow feature at $\vec{k} = (0,0)$. Upon close inspection, such features are seen in 
Fig. \ref{Akom.fig}. The quasiparticle dispersion  relation is at best described in terms of partial 
Kondo screening  of the impurity spins. The remnant non-screened moment  orders antiferromagnetically 
\cite{Zhang00b,Jurecka01b}.

\begin{figure}[]
\begin{center}
\includegraphics[width=.45\textwidth]{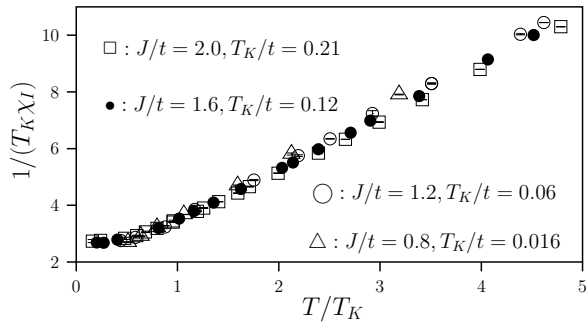} \\
\end{center}
\caption[]{ Impurity spin susceptibility, 
$ \chi_I = \int_{0}^{\beta} {\rm d} \tau \langle m_z^f (\tau) m_z^f (0)  \rangle $, for the Kondo model as
determined with a standard implementation of the Hirsch-Fye algorithm.  From the collapse of the data, 
we can estimate the Kondo temperature for the listed values of $J/t$.}
\label{TK.fig}
\end{figure}

Our major concern is the quasiparticle residue at $\vec{k} = (\pi,\pi)$.  
In Fig. \ref{Scales.fig} we plot this quantity as a 
function of $J/t$ for $ 8 \times 8$ and $12 \times 12$ lattices.  At small values 
of $J/t$ size effects become more and  more important. For $J/t = 0.4$ and $0.8$ we have extrapolated 
the data to  infinite sizes by fitting the $L=4,6,8,12$ QMC results to the form $a + b/L + c/L^2$.  
To compare this scale to the single impurity Kondo scale, we have carried out simulations  of 
the Kondo model:
\begin{equation}
	H = \sum_{\vec{k}, \sigma} \epsilon(\vec{k}) c^{\dagger}_{\vec{k},\sigma} c^{}_{\vec{k},\sigma}  
 + J \vec{S}^{c}_{I} \vec{S}^{f}_{I} 
\end{equation}
with a standard implementation of the Hirsch-Fye  QMC algorithm \cite{HirschFye86}. 
In this algorithm the CPU time scales as 
$\beta^3 N^0$ where $N$ corresponds to the number of 
lattice sites and $\beta$ to the inverse temperature. This scaling behavior allows one  to carry out  
simulations  on arbitrarily large lattices but limits the accessible temperature range. Form the data 
collapse of the 
impurity spin susceptibility 
(see Fig. \ref{TK.fig}) we can extract the Kondo temperature for our given band structure.   Note that 
the so obtained Kondo temperature (triangles in Fig \ref{Scales.fig}) compares remarkably well with 
the mean-field value (solid line in
 Fig. \ref{Scales.fig}).  At small values of  $J/t$ ($J/t =0.4$)  the Kondo temperature is too 
small to compute  with the Hirsch-Fye algorithm and we have to  rely on  the mean-field result 
to compare with the  lattice QMC results.

The data of Fig. \ref{Scales.fig} show several features. \\
i)  The major observation is that within  the considered range of coupling constants the QMC results 
are consistent with 
\begin{equation}
\label{ZproptoTK}
	Z_{(\pi,\pi)} \propto T_K.
\end{equation} 
\\
ii) Due to the occurrence  of antiferromagnetic order below $J_c/t = 1.45  \pm 0.05$, the quasiparticle 
gap tracks $J$.  Even in the presence of this {\it large} quasiparticle gap, the Kondo-like features in 
the single-hole spectral function 
-- flat band around the $(\pi,  \pi)$ point with $Z_{(\pi,\pi)} \propto T_K$ --  survive. \\
iii)  Within our resolution and limitation in lattice size,  we observe no anomaly in the 
quasiparticle residue across the magnetic order-disorder phase transition ($J_c/t \simeq 1.45)$.

{\it  Discussion}
Under the assumption of a rigid band, the  result of Eq. \ref{ZproptoTK}  has important implications. 
Since the quasiparticle gap is determined by the $( \pm \pi, \pm \pi)$ points in the Brillouin zone, 
infinitesimal hole doping away from half-filling yields  a metallic state with Fermi surface 
consisting of hole  pockets centered around the $(\pm \pi, \pm \pi)$ $\vec{k}-$points.  
This leads to a Fermi surface with large Luttinger 
volume  containing both  conduction  and impurity electrons. The effective mass of this  Fermi liquid at
infinitesimal hole dopings is  inversely proportional to the quasiparticle residue: 
\begin{equation}
\label{mstar}
m^{\star}  \propto  \frac{1} {Z_{(\pi,\pi)}}.
\end{equation}
The coherence temperature, $T_{\rm coh}$, is inversely proportional to the effective mass. Hence
Eq. \ref{mstar} along with our QMC result of Eq. \ref{ZproptoTK}  is equivalent to 
\begin{equation}
\label{Conc}
T_{\rm coh} \propto  T_K.
\end{equation}
Let us comment on the implicit assumptions lying behind the above equation.   \\ 
i) Support for the rigid band assumption  follows from the fact the Kondo insulator  is 
adiabatically linked to  band insulator realized in the non-interacting periodic Anderson model. 
This stands in contrast to the Mott insulting state. \\
ii)  That the effective mass is inversely proportional to to quasiparticle residue implicitly 
assumes that the enhancement of $m^{\star}$ is driven by the frequency dependence of the 
self-energy.  We can obtain support for this statement  by computing the effective mass 
as the inverse curvature of the dispersion relation around  $\vec{k} = (\pi,\pi)$ thus capturing the 
mass enhancement origination from the frequency as well as momentum dependence of the self-energy. 
Within our accuracy, this quantity tracks $1/Z_{(\pi,\pi)}$  as a function of $J/t$ thus confirming 
the validity of our assumption.

In conclusion, we have carried out  high precision QMC calculations  of the half-filled  KLM on a
square lattice. We have shown that from weak to strong coupling and across the magnetic quantum phase 
transition, a doped hole has momentum $(\pi,\pi)$ and quasiparticle residue which
tracks the Kondo scale of the corresponding single impurity model. Assuming a rigid band, this 
result leads to the conclusion that the $J$-dependence of coherence scale in the metallic state at 
small dopings away from half-filling tracks the Kondo scale.   
Since this result compares favorably  
with large-N and  dynamical mean-field theories  which omit spatial fluctuations, 
we can follow the idea that the coherence scale is insensitive to the RKKY magnetic scale.

Acknowledgments.  The calculations were carried out on the Hitachi SR8000 of the Leibniz Rechenzentrum 
(LRZ) Munich.  We thank this institution for generous allocation of CPU time.  Stimulating discussions 
with I. Milat and O. Shushkov are equally acknowledge.

%\bibliographystyle{./prsty}
%\bibliography{/home/assaad/TeX/fassaad}

\end{document}